\documentclass{article}
\usepackage{spconf,amsmath,amssymb,graphicx}
\usepackage{xcolor,colortbl}
\usepackage{makecell}
\usepackage{pdfpages}
\usepackage[]{algorithm2e}
\usepackage{todonotes}
\usepackage{float}
\usepackage{calc}
\usepackage{graphicx}
\usepackage{color}
\usepackage{subcaption}
\usepackage{balance}
\usepackage[final]{changes}
\usepackage{pgfplots}
\usepackage[hyphens]{url}
\usepackage{changes}
\usepackage{hyperref}
\usepackage{multirow}
\usepackage{bm}

\ninept

\newcommand{\Dece}{D_{\text{ECE}}}
\newcommand{\Cllrmin}{C_{\text{llr}}^{\text{min}}}

\title{Hiding speaker's sex in speech using zero-evidence\\speaker representation in an analysis/synthesis pipeline}
\name{Paul-Gauthier Noé$^1$, Xiaoxiao Miao$^2$, Xin Wang$^2$}
\secondlinename{Junichi Yamagishi$^2$, Jean-François Bonastre$^1$, Driss Matrouf$^1$\thanks{This work was done when Paul-Gauthier Noé was visiting Yamagishi Laboratory, National Institute of Informatics, Japan. It was supported by the VoicePersonae project ANR-18-JSTS-0001, JST CREST Grants (JPMJCR18A6 and JPMJCR20D3), MEXT KAKENHI Grants (21K17775, 21H04906, 22K21319), and Google AI for Japan program.}
}
\address{$^1$Laboratoire Informatique d'Avignon, Avignon Université, France\\
  $^2$National Institute of Informatics, Japan}

\begin{document}
\maketitle

\begin{abstract}
The use of modern vocoders in an analysis/synthesis pipeline allows us to investigate high-quality voice conversion that can be used for privacy purposes. Here, we propose to transform the speaker embedding and the pitch in order to hide the sex of the speaker. ECAPA-TDNN-based speaker representation fed into a HiFiGAN vocoder is protected using a neural-discriminant analysis approach, which is consistent with the zero-evidence concept of privacy. This approach significantly reduces the information in speech related to the speaker's sex while preserving speech content and some consistency in the resulting protected voices.
\end{abstract}
\section{Introduction}
\label{sec:intro}
Privacy considerations are rising in speech technology research~\cite{NAUTSCH2019441}. The VoicePrivacy challenge~\cite{Tomashenko2020} focuses on hiding the full identity of the speaker. However, there might be situations where the user requires the protection of only one or a few of their personal attributes, for instance, their sex, native language, emotional or health state. This approach to privacy is known as \emph{user configurable}~\cite{aloufi2020privacy} and \emph{attribute driven} privacy~\cite{NOE2022}. In this work, we focus on sex as an attribute to hide in speech signals.

Recently, methods have been proposed for this purpose. In~\cite{genderneut}, in order to avoid sex-related bias in speech model training, the authors proposed making speech \emph{sex-neutral} beforehand by automatically searching for pitch and formants shifting that would lead to the maximum uncertainty sex classifier score,~i.e.,\ 50\%. However, there is no guarantee that the classifier they used is well calibrated~\cite{pmlr-v70-guo17a}, and a search for shifting parameters must be done for each utterance. Here, we prefer to have a single transformation that can be applied regardless of the input utterance, which appears to us more suitable for a real-life application of privacy systems.
In~\cite{stoidis22_interspeech}, the authors proposed removing the speaker's sex using an adversarial approach. Their approach also aims to protect the speaker's identity instead of leaving the speaker's other information unchanged.

Here, we want to alter only the speaker's sex while preserving the other speaker-related variabilities. We therefore consider the explicit disentanglement of sex information as a desirable step. In~\cite{NOE2022}, we proposed a neural-discriminant-analysis-based approach for disentanglement. The sex variable is represented as a log-likelihood-ratio (LLR) that can be set to zero for privacy, which is consistent with the zero-evidence recognition framework~\cite{Nautsch2020} and Shannon's perfect secrecy~\cite{shannon}. However, this approach has been designed for vector inputs and applied to speaker embeddings. Extending it to waveforms is challenging, and we therefore want, as a first step, to include it in an analysis/synthesis framework for voice conversion for sex protection. We use the HiFiGAN vocoder~\cite{hifigan} fed by the $f_0$ trajectory, a HuBERT soft content representation~\cite{van2022comparison}, and an ECAPA-TDNN~\cite{desplanques2020ecapa} speaker embedding. Once the analysis/synthesis pipeline is trained, we apply the protection proposed in~\cite{NOE2022} to the speaker vector and an affine transformation to the $f_0$ to remove sex-related information.
In our experiments, we test the protection ability of our approach in terms of sex recognition performance with both \emph{ignorant} and \emph{semi-informed} attacks~\cite{srivastava2019evaluating}, respectively considered \emph{weak} and \emph{strong} attacks. We evaluate the performance of automatic speech recognition (ASR) and speaker verification (ASV) as downstream tasks. Listening tests are also done in order to assess human ear perception of protected speech.

Randomly assigning a target sex to each speaker could lead to better protection results using our evaluation protocol. However, we want to inform the reader that our concept of privacy is not to fool the attacker but rather to not provide any evidence about speaker's sex, resulting in some kind of sex-neutral voice\footnote{Audio samples and model are available at \url{https://github.com/nii-yamagishilab/speaker_sex_attribute_privacy}}.

\section{Attribute privacy and zero-evidence}

Most of the approaches in voice conversion for privacy aim to hide the full speaker identity~\cite{NAUTSCH2019441,Tomashenko2020}. Attribute privacy aims instead to hide only one or a few attributes of the speaker~\cite{NOE2022}, making it possible to look for a better compromise between utility and \emph{user configurable} privacy~\cite{aloufi2020privacy}. The attributes can be personal information such as the speaker's sex, emotional and health state and so on. The attacker's knowledge on an attribute he or she wants to infer is represented by a discrete probability distribution over the possible outcomes (male and female in our case). The Bayes' rule provides a natural way to update the attacker's belief in light of observed data. For perfect secrecy/privacy, posterior and prior knowledge has to be the same~\cite{shannon}, which corresponds to a LLR equal to zero; this is \emph{zero-evidence}~\cite{Nautsch2020}.

In attribute privacy, the attributes to conceal have a relatively low number of possible outcomes. For instance, if the speaker appears to be a French native and an attacker wants to infer from which region the speaker comes from, the number of possible outcomes is 18,~i.e.,\ the number of administrative regions in France. In this case, where the number of classes is significantly lower than the dimensionality of the data, the latter can be transformed such that a group of components embeds the attribute-related variability, while the other components contain the residual variability. Once this separation has been done, the attribute variability can be annihilated for privacy. This approach differs from common privacy tasks where the number of classes to make indistinguishable can be arbitrarily large.

In~\cite{NOE2022}, we proposed a nonlinear discriminant analysis that allows for manipulating the LLR related to the sex attribute in speaker representation. For privacy, the LLR can be set to zero, which is consistent with \emph{zero-evidence}. However, in its current design, this approach cannot be applied to raw speech data because speech is time dynamical. In the next section, we propose a way to get around this problem by protecting intermediate features in an analysis/synthesis pipeline instead of trying to sanitize raw speech directly.

\section{Proposed protection system}
\label{sec:proposed}

Analysis/synthesis is the process of extracting speech features from which the original speech signal can be recovered using a vocoder. In speech technology, this approach has been widely used. The intermediate characterisation of speech can be used for speech transmission, voice conversion, speech anonymisation... In speech anonymisation, we want a part of the intermediate features to represent speaker-related information that can be manipulated for privacy. In~\cite{miao22_odyssey}, the authors proposed updating the first VoicePrivacy's baseline~\cite{Tomashenko2020} by replacing the neural source-filter vocoder~\cite{wang19_ssw} with a HiFiGAN vocoder~\cite{hifigan}. They also replaced the Kaldi TDNN speaker embedding~\cite{snyder2018x} (xvector) with an ECAPA-TDNN speaker embedding~\cite{desplanques2020ecapa} considered to be the state-of-the-art representation for ASV. They finally got rid of the acoustic model by using instead a HuBERT-based soft content representation~\cite{van2022comparison}. In~\cite{miao22_odyssey}, they used this system for the VoicePrivacy task and studied its application to unseen languages. In this paper, we use this system, but we replace the speaker embedding averaging used for voice anonymisation with the discriminant-analysis-based protection in~\cite{NOE2022} and add an affine transformation of the $f_0$ trajectory for sex protection.\\
\noindent
\textbf{Speaker representation protection:} In~\cite{miao22_odyssey,Tomashenko2020}, for anonymisation, the original xvector of an utterance is replaced with an average of xvectors randomly selected from a pool of speakers. Here, we want to conceal the sex of the speaker only. We propose using the discriminant-analysis-based approach presented in~\cite{NOE2022} and discussed above. We recall here in more detail how this can be used for the concealment of sex-related information in speaker embeddings.
The idea is to use normalizing-flow neural-transformation~\cite{dinh2017} to learn an invertible mapping from the speaker embedding space to a base space where the class-conditional densities are carefully chosen such that only the first component embeds the sex information in the form of a LLR {\small$\log \frac{P(x|C=0)}{P(x|C=1)}$} (where $C$ is for class, 0 for male, 1 for female). When the LLR is zero, the observation $x$ is equally likely to come from both classes, resulting in no change in the belief of the observer/attacker. Therefore, for protection, the observed embedding is mapped into the base space where the first dimension (LLR) is set to zero before mapping back to the observation space.\\
\noindent
\textbf{$\bm{f_0}$ protection:} The fundamental frequency ($f_0$) is known to contain information about the sex. We therefore apply an affine transformation to the $f_0$ to force a fixed target $f_0$ trajectory mean and standard deviation that we expect to be sex-neutral. They are computed from a training set where the means and standard deviations from $f_0$ trajectories are first averaged at the speaker level and are then averaged over all males and all females resulting in two means and two standard deviations $f_0$ (one for each sex). Then, the target mean $f_0$ is obtained by taking the average between the male and the female mean $f_0$, and the target standard deviation is obtained by taking the average between the male and the female standard deviation. This careful averaging is done in order to avoid bias due to an unbalanced number of utterances per speaker and speakers per sex.

\begin{figure}
    \centering
    \includegraphics[width=0.44\textwidth]{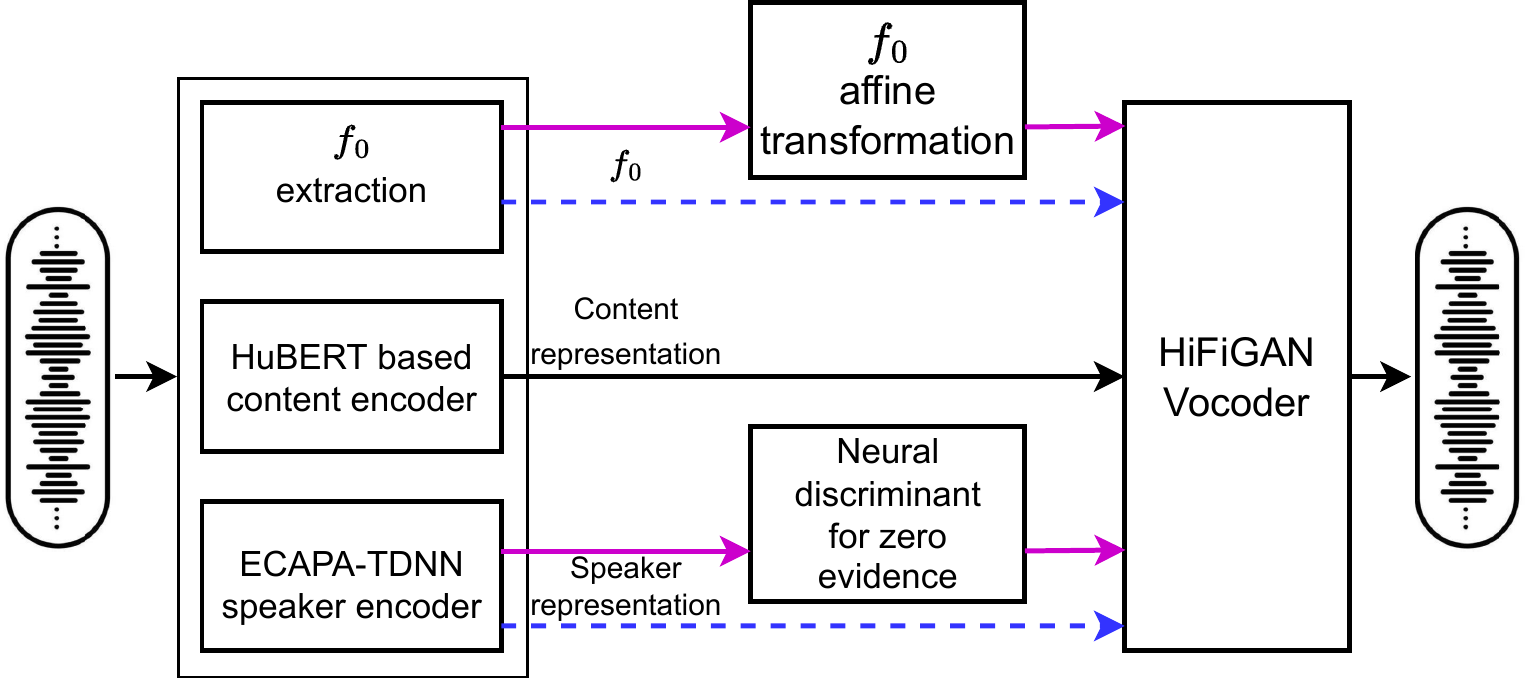}
    \caption{Architecture of our system. Blue-dashed path is used during training and purple one during protection.}
    \label{fig:diagram}
\end{figure}

Figure~\ref{fig:diagram} shows an outline of our system. The blue-dashed arrows show the training path of the HiFiGAN. The feature extractors are pretrained and fixed. Once the vocoder has been trained, the purple path is used. Both the $f_0$ and the speaker representation are transformed to reduce the sex-related information they contain. The content representation is assumed to not contain sex-related information. However, in real applications, the speaker might explicitly reveal their sex but we do not consider this scenario and instead focus on hiding the sex information in the acoustic features.

\section{Experiments}

This section presents the sets used for training and testing the system, the baselines with which we compare it and experimental results.
\vspace{-0.18cm}
\subsection{Training and testing sets}
\textbf{Vocoder's training:}
LibriTTS train-clean-100~\cite{libritts} was used to train the HiFiGAN as in~\cite{miao22_odyssey}. The feature extraction modules were pretrained and fixed. ECAPA-TDNN~\cite{desplanques2020ecapa} with 80-coefficient FBank features~\cite{miao22_odyssey} was used for the 192-dimensional speaker representation extraction and was trained on VoxCeleb2 development set~\cite{voxceleb2}. The 200-dimensional content representation was extracted by a HuBERT soft content encoder \cite{van2022comparison} fine-tuned from a pretrained HuBERT base model\footnote{\url{https://github.com/pytorch/fairseq/tree/main/examples/hubert}} on LibriTTS train-clean-100. 
Its training procedure is detailed in~\cite{miao22_odyssey}.
We used YAAPT~\cite{yaapt} for the $f_0$ extraction, which does not require any training.\\
\noindent
\textbf{Training of protection modules:}
Once the analysis/synthesis system is trained, it can be used for privacy by manipulating the speaker representation and the $f_0$. In our experiments, the former was protected using the discriminant analysis for zero-evidence sex recognition presented in~\cite{NOE2022} and summarised in Section~\ref{sec:proposed}. It was trained on ECAPA-TDNN speaker embeddings~\cite{desplanques2020ecapa} from LibriTTS train-other-500. The target $f_0$ mean and standard deviation were computed as described in~\ref{sec:proposed} from LibriTTS train-other-500 also.\\
\noindent
\textbf{Testing sets:}
The VoicePrivacy challenge provides a complete evaluation protocol. For conciseness, we merged its libri\_dev and libri\_test sets to assess our system, resulting in 35 females with a total of 1185 utterances and 34 males with a total of 1136 utterances.

\subsection{Baselines}

We compared the proposed approach with two baselines. The first, called \emph{global}, is the same as our proposed approach but instead of using the neural-discriminant-analysis-based protection of the speaker representation, we simply fed the HiFiGAN with the same global averaged xvector for all utterances. The averaging was done in such a way as to avoid bias as it was done for the computation of the target $f_0$ trajectory moments. In this case, we expect that the sex of the original speaker will be hidden but that all the speaker information will be altered such that the resulting voices all look the same. The second baseline transforms only the $f_0$ using time domain pitch synchronous overlap add (TDPSOLA)~\cite{MOULINES1990453} where, because we know that sex information is also contained in the spectral envelope, we expect that the sex of the speaker will not be satisfactorily hidden.

\subsection{Results}

\begin{figure}
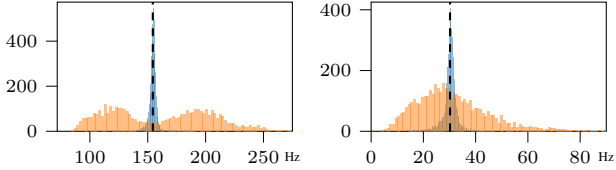

    \centering
    \begin{subfigure}[t]{0.23\textwidth}
        \input{figures/histogram_f0/meanf0hist.tex}
    \end{subfigure}
    \begin{subfigure}[t]{0.23\textwidth}
        \input{figures/histogram_f0/stdf0hist.tex}
    \end{subfigure}
    \caption{Mean $f_0$ (left) and standard deviation $f_0$ (right) histograms for original speech (orange) and for generated protected speech (blue) with target $f_0$ mean and standard deviation (dashed lines).}
    \label{fig:histf0}
\end{figure}

We report the results for \emph{original} speech, \emph{synthesised},~i.e.,\ fed into our system but without xvector and $f_0$ transformations, protected with the \emph{proposed} approach,~i.e.,\ with xvector and $f_0$ transformations as presented in Section~\ref{sec:proposed}, together with the \emph{global} approach and the \emph{TDPSOLA} approach. To assess the protection, we report to which extent an automatic sex classifier is able to detect the sex of the speaker. We then present results for ASR and ASV as downstream tasks and voice similarity matrices. Finally, we present listening test results. First, Figure~\ref{fig:histf0} shows histograms of generated $f_0$ trajectory mean and standard deviation when the proposed approach was used. We can see that the mean and standard deviation of the generated $f_0$ follow the target ones. Indeed, their histograms (in blue) are narrow around the target values shown by the dashed lines.

\subsubsection{Protection assessment: Automatic sex classification}

To objectively assess the protection performance of the systems, we report the results of automatic sex recognition. This section is concerned with automatic attacks only. An attacker may try to infer the sex of the speaker by listening manually to the data. This point will be discussed in Section~\ref{sec:listen}.
We propose two kinds of attack: one where the classifier is trained on original speech data and one on protected speech data.
The former corresponds to an \emph{ignorant} attacker and the latter is analogous to a \emph{semi-informed} attacker~\cite{srivastava2019evaluating}. The \emph{ignorant} attacker does not have access to the protection system or may not be aware that the data has been protected. In this case, it uses a sex classifier trained on natural non-protected speech. The \emph{semi-informed} one is the strongest attack we consider. In this case, the attacker has access to the protection system. He or she can apply it to data he or she will be using for training the automatic sex classifier. The resulting classifier therefore benefits from the sex-related information that could remain in the protected data.
The classifier we used in our experiments is based on fine-tuned HuBERT base features extraction (with frozen convolution), statistical pooling and multilayer perceptron. Table~\ref{tab:attackmosasv} reports the results in terms of two metrics: the equal error rate (EER) and the $\Dece$~\cite{Nautsch2020}. The latter is a positive measure of the expected amount of information disclosed to the attacker when observing the output of the classifier. For privacy, we want a low $\Dece$. The first line shows the initial ability to distinguish the sex of the speakers. We can see from the second line that this is slightly altered when processed even without protection applied. The next two lines show how the classification performance drops when protection is applied. For the ignorant attack, we have a drop in $\Dece$ of 78\% for the \emph{proposed} approach and 66\% for \emph{global}. The methods are also robust to the semi-informed attack with a drop in $\Dece$ of 65\% and 60\%. The \emph{TDPSOLA} baseline is not competitive, which is not a surprise because it alters only the $f_0$ while it is known that differences in vocal tract shape between males and females are significantly related to the spectral envelope. However, we do not have a clear explanation as to why the \emph{proposed} approach protects better than \emph{global} baseline does. This could be due to uncontrolled bias in the data but this requires further study.
\begin{table}
    \centering
    \caption{Sex classification results for protection assessment, and automatic speech recognition WER.}
    \resizebox{0.48\textwidth}{!}{
    \begin{tabular}{|c||c|c|c|c||c|}
        \hline
        &\multicolumn{2}{c|}{ignorant} & \multicolumn{2}{c||}{semi-informed} & ASR \\
        \hline
        system & {EER {\scriptsize[\%]}} &  $D_{\text{ECE}}$ {\scriptsize[bit]} & {EER {\scriptsize[\%]}} & $D_{\text{ECE}}$ {\scriptsize[bit]} & WER {\scriptsize[\%]}\\
        \hline
        \hline
        original & 3.67 & 0.578 & \cellcolor{lightgray} & \cellcolor{lightgray} & 4.02 \\
        \hline
        synthesised & 4.32 & 0.542 & 4.01 & 0.593 & 4.79\\
        \hline
        global & 24.95 & 0.198 & 20.60 & 0.233 & 4.92\\
        \hline
        proposed  & 28.99 & 0.128 & 24.13 & 0.200 & 4.81\\
        \hline
        TDPSOLA & 6.30 & 0.504 & 4.36 & 0.542 & 4.43\\
        \hline  

    \end{tabular}
    }
    \label{tab:attackmosasv}
\end{table}
\subsubsection{Automatic speech recognition and speaker verification}
In this section, we want to ensure that automatic speech recognition and automatic speaker verification can still be performed as downstream tasks. We used the same ASR evaluation as in the VoicePrivacy challenge~\cite{Tomashenko2020}. The word error rate (WER) is reported in Table~\ref{tab:attackmosasv}. Processing the speech increases the WER slightly, but among the two systems that provide good protection, the proposed approach seems to alter the ASR performance less. At worst, 0.9\% is added to the WER which is a relatively low price to pay for privacy.

\begin{figure}
    \begin{subfigure}[b]{0.23\textwidth}
        \centering
        \includegraphics[width=0.8\textwidth]{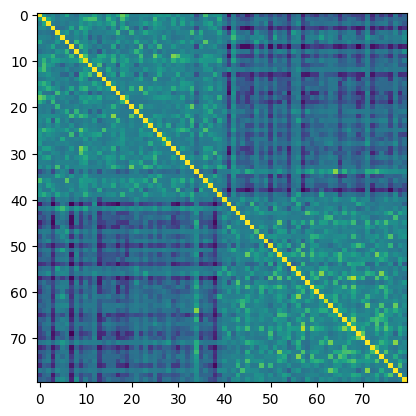}
        \caption[]
        {{\small original}}  
    \end{subfigure}
    \hfill
    \begin{subfigure}[b]{0.23\textwidth}  
        \centering 
        \includegraphics[width=0.8\textwidth]{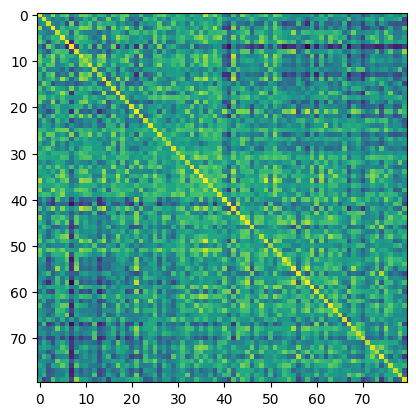}
        \caption[]%
        {{\small protected with \emph{proposed}}}   
    \end{subfigure}
    \vskip\baselineskip
    \vspace{-0.4cm}
    \begin{subfigure}[b]{0.23\textwidth}   
        \centering 
        \includegraphics[width=0.8\textwidth]{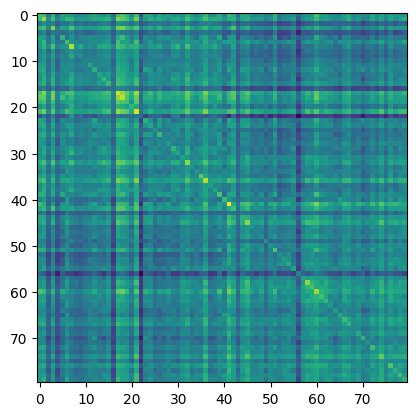}
        \caption[]
        {{\small protected with \emph{global}}}
    \end{subfigure}
    \hfill
    \vspace{-0.2cm}
    \begin{subfigure}[b]{0.23\textwidth}   
        \centering 
        \includegraphics[width=0.8\textwidth]{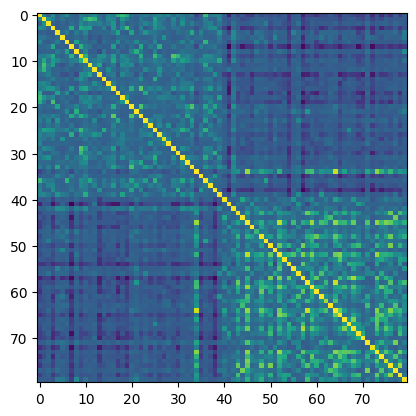}
        \caption[]
        {{\small protected with \emph{TDPSOLA}}}   
    \end{subfigure}
    \caption{Voice log-similarity matrices~\cite{NOE2022101299}.}
    \label{fig:matrices}
\end{figure}

We also report ASV results and voice similarities matrices to check if, after protection, ASV can still perform. We used the same ASV system used for evaluation in the VoicePrivacy challenge. It consists of a Kaldi TDNN speaker embedding extractor~\cite{snyder2018x} with a PLDA backend. Both enrolment and test utterances were processed by the system.
The EER and $\Cllrmin$~\cite{brummer2006application} are reported in Table~\ref{tab:asv}. We can see that processing the data without protection already slightly reduces the ASV performance. This suggests that the HiFiGAN vocoder results in a small distortion or domain shift. However, applying protection further reduces the ASV performance. For \emph{global}, all the speaker variability in the xvector is annihilated by the global averaging, therefore increasing the confusion between voices. With the \emph{proposed} approach, the xvector is disentangled in order to alter only the speaker's sex. Other speaker variabilities are preserved and, as expected, the protected voices remain consistent to some extent. Indeed, the proposed approach does far better than the global one although, compared with original data, significant ASV ability is lost with an increase in $\Cllrmin$ from 0.278 to 0.445 and from 0.040 to 0.345 for female and male respectively. In addition to the domain shift induced by the HiFiGAN synthesis, this drop in performance could be explained by both the reduction in sex information as a component of the speaker variability that helps in distinguishing speakers, not only a male speaker from a female one but also between speakers with the same sex, and also by imperfect disentanglement of the sex component from other speaker-related information.
\begin{table}[]
    \centering
    \caption{Automatic speaker verification results. F and M refer respectively to in-between female and in-between male trials, while FM refers to cross-sex trials.}
    \resizebox{0.44\textwidth}{!}{
    \begin{tabular}{|c|c|c|c|c|c|c|}
    \hline
    \multirow{2}{*}{system}& \multicolumn{3}{c|}{EER {\scriptsize[\%]}} & \multicolumn{3}{c|}{$\Cllrmin$ {\scriptsize[bit]}}\\
    \cline{2-7}
    & F & M & FM & F & M & FM \\
    \hline
    \hline
    original & 8.15 & 1.13 & 5.77 & 0.278 & 0.040 & 0.204\\
    \hline
    synthesised & 9.42 & 7.18 & 6.86 & 0.325 & 0.245 & 0.240\\
    \hline
    global & 39.88 & 39.81 & 35.86 & 0.931 & 0.943 & 0.903 \\
    \hline
    proposed & 13.22 & 9.85 & 11.55 & 0.445 & 0.345 & 0.407 \\
    \hline
    TDPSOLA & 9.36 & 1.26 & 6.38 & 0.332 & 0.046 & 0.237 \\
    \hline
    \end{tabular}
    }
    \label{tab:asv}
\end{table}
\noindent
In~\cite{NOE2022101299,noe20_interspeech}, the authors proposed, in the context of the VoicePrivacy initiative, to visualise speaker voice similarity matrices to investigate the behavior of a protection system at both a speaker and global level. Here, our task is different, but we can still visualise voice similarity matrices to assess the consistency of the protected voices and to pay attention to any sex-related patterns that could appear in the matrices. We report four of these matrices in Figure~\ref{fig:matrices}. Speakers were grouped by sex such that squares appear in the matrix (a). Indeed, male speakers generally look more like other males than females and vice versa. When we have good sex protection in (b) and (c), we can see that these squares tend to disappear. The near disappearance of the diagonal in (c) confirms that \emph{global} is not suitable enough to preserve other speaker variabilities compared with \emph{proposed}.
\vspace{-0.05cm}
\subsubsection{Listening tests}
\label{sec:listen}
\begin{figure}
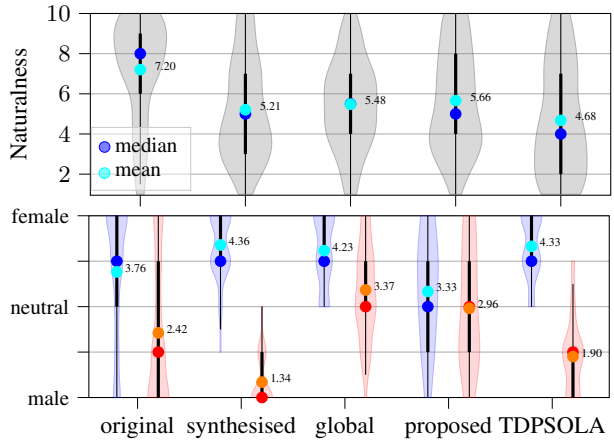

    \centering
    \begin{subfigure}[t]{0.5\textwidth}
        \input{figures/listening_test/naturalness}
    \end{subfigure}
    \vskip\baselineskip\vspace{-0.7cm}
    \begin{subfigure}[t]{0.5\textwidth}
        \input{figures/listening_test/genderscores}
    \end{subfigure}
    \caption{Listening test results. Violin plots of perceived speech naturalness (top). Violin plots of perceived speaker's sex (bottom), blue for female and red for male; blue and red dots show medians, cyan and orange dots show means.}
    \label{fig:listen}
\end{figure}

The results presented so far show the machine's perception. In this section, we discuss how the human ear perceives protected speech by reporting listening test results. 19 listeners, all native English speakers, were asked to assess the naturalness of speech on a discrete scale from 1 (unnatural) to 10 (natural) and whether the speech sounded like a male (1), a  female (5), or neutral (3), also allowing for some nuance with scores of 2 and 4. \emph{Neutral} refers here to the zero-evidence formulation of privacy where we want the data to provide no evidence about the speaker's sex such that the listener posterior belief remains equal to the prior one. Figure~\ref{fig:listen} shows the naturalness and sex perception scores. The speech processed by the analysis/synthesis even without protection does not sound as natural as the original speech. However, applying the protection does not further decrease the naturalness. As expected, \emph{TDPSOLA} does not sufficiently change the perception of the sex. While \emph{global} seems to change the perception of the speech from males, it does not have the expected behavior for females\footnote{Again, we do not have a clear explanation as to why \emph{global} does not protect the data as well as \emph{proposed}.}. The \emph{proposed} approach works for both male and female with a good average score close to 3 which tends to make attacks by listening inefficient. However, for better zero-evidence protection of each utterance, it would have been better to have narrower distributions around the neutral score.
\section{Conclusion}
For privacy reasons, this paper proposed removing the sex of the speaker in speech using an analysis/synthesis-based voice conversion pipeline. An affine transformation is applied to the pitch, and the speaker representation is disentangled using a neural-discriminant analysis in order to conceal the speaker's sex-related information. The latter is consistent with the zero-evidence framework. The protection ability of the system was checked by means of an automatic sex classifier considering both an \emph{ignorant} and a \emph{semi-informed} attacker. Automatic speech recognition can still be applied on protected speech. Although the automatic speaker verification is deteriorated, the protected voices remain consistent to a certain extent. A listening test showed that the naturalness of the protected speech is satisfactory and the perception of the speaker's sex is altered, making attacks by listening more difficult.

In the future, we are interested in extending this work to other attributes like, for instance, accents. However, as accents involve more classes and are rarely labeled in large datasets, we expect that handling them will be even more challenging.

\newpage
\bibliographystyle{IEEEbib}
\balance\bibliography{bib}

\end{document}